\documentclass[preprint2]{emulateapj}
\usepackage{graphicx}
\usepackage{natbib}
\def\solmasp{$\mathrm{M_\odot}$}

\newcommand{\thebrick}{G0.253+0.016}
\newcommand{\uvrad}{I$_{\rm ISRF}$}
\newcommand{\crrate}{I$_{\rm CR}$}
\newcommand{\habing}{G$_0$}
\newcommand{\crlocal}{I$_{\rm CR, 0}$}

\begin{document}

\title{On the temperature structure of the Galactic Centre cloud \thebrick}

\author
{Paul C.\ Clark$^1$, Simon C.O. Glover$^1$, Sarah E. Ragan$^2$, Rahul Shetty$^1$ \& Ralf S. Klessen$^1$}

\affil{$^1$Universit\"at Heidelberg, Zentrum f\"ur Astronomie, Institut f\"ur Theoretische Astrophysik, Albert-Ueberle-Str.\ 2, 69120 Heidelberg, Germany. \\
$^2$Max Planck Institut f\"ur Astronomie, K\"onigstuhl 17, 69117 Heidelberg, Germany.
\break email: p.clark@uni-heidelberg.de, glover@uni-heidelberg.de, ragan@mpia.de, r.shetty@uni-heidelberg.de, klessen@uni-heidelberg.de}

\begin{abstract}

We present a series of smoothed particle hydrodynamical models of \thebrick~(also known as ``The Brick''), a very dense molecular cloud that lies close to the Galactic Centre.
We explore how its gas and dust temperatures react as we vary the strength of both the interstellar radiation field (ISRF) and the cosmic ray ionisation rate (CRIR). The cloud has an 
extent in the plane of the sky of roughly 3.4~pc  $\times$ 9.4~pc. As its size along the line-of-sight is unknown, we consider two cases. In our fiducial, high-density model, we adopt a depth along the line-of-sight of 3.4 pc, and in the low-density model, we assume an extent along the line-of-sight of 17 pc. To recover the observed gas and dust temperatures, we find that the ISRF must be around $1000\,$ times the solar neighbourhood value, and the CRIR must be roughly $10^{-14} \rm \, s^{-1}$, regardless of the geometries studied. For such high values of the CRIR, we find that  cooling in the cloud's interior is dominated by neutral oxygen, in contrast to standard molecular clouds, which at the same densities are mainly cooled via CO. Our results suggest that the conditions near \thebrick~are more extreme than those generally accepted for the inner 500 pc of the galaxy.

\end{abstract}

\maketitle

\section{Introduction}
\label{sec:intro}

\begin{deluxetable*}{c  c  c  c  c  c  c  c  c  c  c c}
\tablecaption{\label{table:sims} Summary of the simulations}
\tablehead{
\colhead{Model}  & \colhead{$^{2}$L$_x$} & \colhead{$^{3}$L$_y$} & \colhead{$^{4}$L$_z$} & \colhead{$^{5}$$\Sigma_{\rm min, 0}$}  & \colhead{$^{6}$$n_{\rm 0}$} & \colhead{$^{7}$\uvrad} & \colhead{$^{8}$\crrate} & \colhead{$^{9}$$x(\rm H_2)$} & \colhead{$^{10}$$x(\rm CO)$} & \colhead{$^{11}$$x(\rm C^+)$} & \colhead{$^{12}$$x(\rm O)$} \\
\colhead{} & \colhead{[pc]} & \colhead{[pc]} & \colhead{[pc]} & \colhead{[cm$^{-2}$]} & \colhead{[cm$^{-3}$]} & \colhead{[G$_0$]} & \colhead{[s$^{-1}$]} & \colhead{} & \colhead{} & \colhead{} & \colhead{}
}
\startdata
1 & 9.4 & 3.4 & 3.4  & $3.6 \times 10^{23}$ & $3.5 \times 10^4$ & 1000 & $3\,\times\,10^{-14}$ & 0.477 & $1.5 \times 10^{-5}$ & $1.33 \times 10^{-5}$ & $ 3.06 \times 10^{-4}$ \\
2 & 9.4 & 3.4 & 3.4  & $3.6 \times 10^{23}$ & $3.5 \times 10^4$ & 100  & $3\,\times\,10^{-15}$ & 0.500 & $9.59 \times 10^{-5}$ & $8.02 \times 10^{-6}$ & $2.24 \times 10^{-4}$ \\
3 & 9.4 & 3.4 & 3.4  & $3.6 \times 10^{23}$ & $3.5 \times 10^4$ & 1000 & $3\,\times\,10^{-16}$ & 0.500 & $1.11 \times 10^{-4}$ & $3.08 \times 10^{-6}$ & $2.01 \times 10^{-4}$ \\
4 & 9.4 & 3.4 & 17.0 & $7.3 \times 10^{22}$ & $6.7 \times 10^3$ & 100  & $3\,\times\,10^{-16}$ & 0.500 & $6.65 \times 10^{-5}$ & $1.76 \times 10^{-5}$ & $2.53 \times 10^{-4}$ \\
5 & 9.4 & 3.4 & 17.0 & $7.3 \times 10^{22}$ & $6.7 \times 10^3$ & 100  & $3\,\times\,10^{-15}$ & 0.496 & $2.62 \times 10^{-5}$ & $2.41 \times 10^{-5}$ & $2.94 \times 10^{-4}$  \\
6 & 9.4 & 3.4 & 17.0 & $7.3 \times 10^{22}$ & $6.7 \times 10^3$ & 1000 & $3\,\times\,10^{-16}$ & 0.497 & $6.10 \times 10^{-5}$ & $2.52 \times 10^{-5}$ & $2.59 \times 10^{-4}$  \\
\enddata
\tablecomments{$^{2, 3, 4}$ Initial physical dimensions of the cloud. $^5$ Minimum column density, measured along the shortest axis.  $^6$ Initial hydrogen nuclei
number density. $^7$ Strength of the interstellar radiation field, in units of the local value. $^8$ Cosmic ray ionisation rate. $^{9, 10, 11, 12}$ Final fractional chemical abundances in the cloud, measured at the point at which the first core goes into runaway collapse. These are quoted with respect to the number of H nuclei. A fully molecular gas therefore
has $x(\rm H_2) = 0.5$. The total carbon and oxygen abundances in the models are $1.4 \times 10^{-4}$ and $3.2 \times 10^{-4}$ respectively.}
\end{deluxetable*}

The environmental conditions in the Galactic Centre (GC) provide an extreme test of our current understanding of the star formation process (e.g.\ \citealt{padelis2010, kdm2012, Longmore2013,kru13}). With both stronger background radiation fields and higher cosmic-ray fluxes compared to clouds in the solar neighborhood, star formation is predicted to occur at higher volume and column densities than is typical in a standard giant molecular cloud \citep{ekw2008}. 

One notable example is \thebrick~(also referred to as the ``The Brick''), which displays both extremely high column and volume densities, yet very little sign of star formation \citep{guesten1981, Lis94, Longmore2012}. Despite the current lack of star formation, the physical conditions in this object are thought to be similar to those required for the formation of massive stellar clusters \citep{Longmore2012}. 

In this paper we investigate the influence of the extreme GC environment on the thermodynamics of dense and massive molecular clouds, in an attempt to better understand the initial conditions for star formation in the inner molecular zone. We adopt values for the interstellar radiation field and the cosmic ray ionisation rate that are significantly higher than those measured in solar neighborhood molecular clouds. For more fundamental parameters such as the mass, dimensions, and turbulent velocity dispersion of the clouds, we take the values for \thebrick~reported by \citet{Longmore2012}. In contrast to the other clouds in the GC, the apparent lack of star formation in G0.253+0.016 makes it an ideal candidate for studying the effects of the environmental conditions on the thermal balance of the cloud.

\begin{figure}[t]
\includegraphics[width=3.5in]{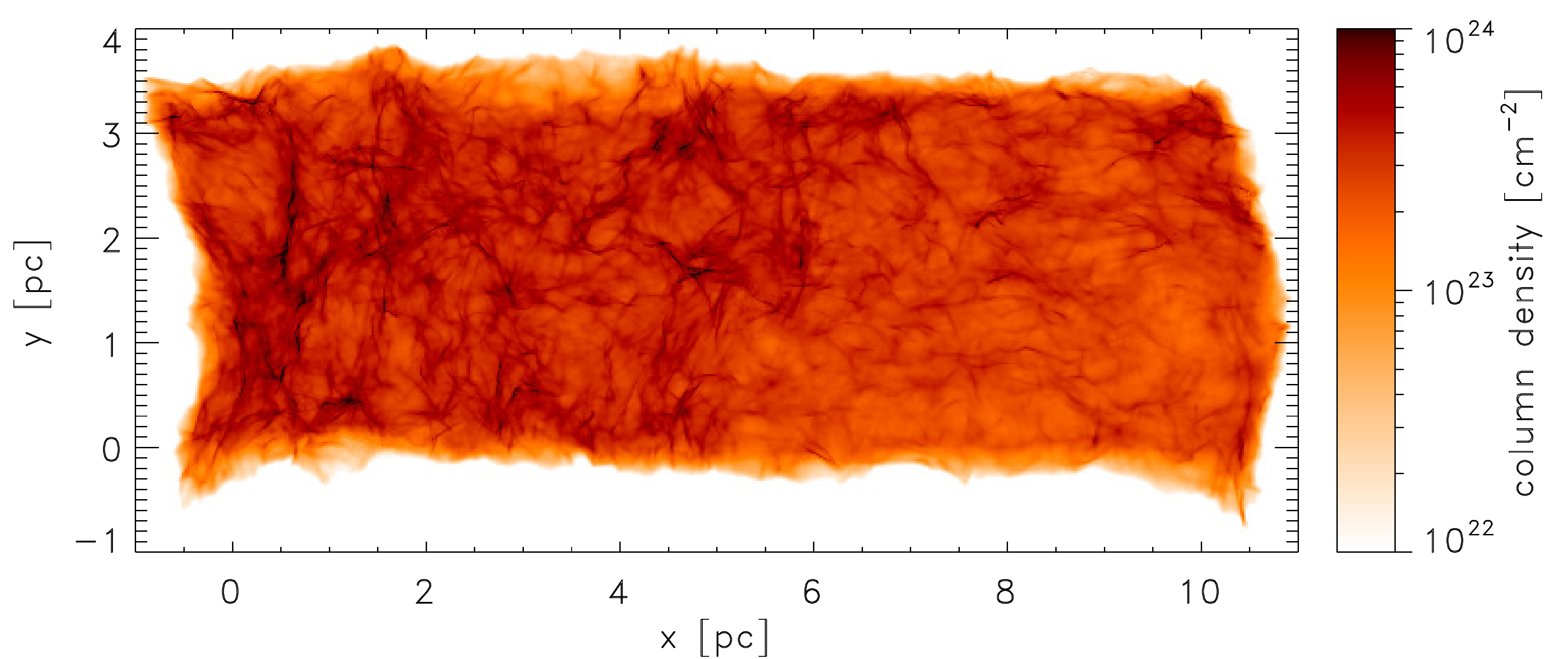}
\includegraphics[width=3.5in]{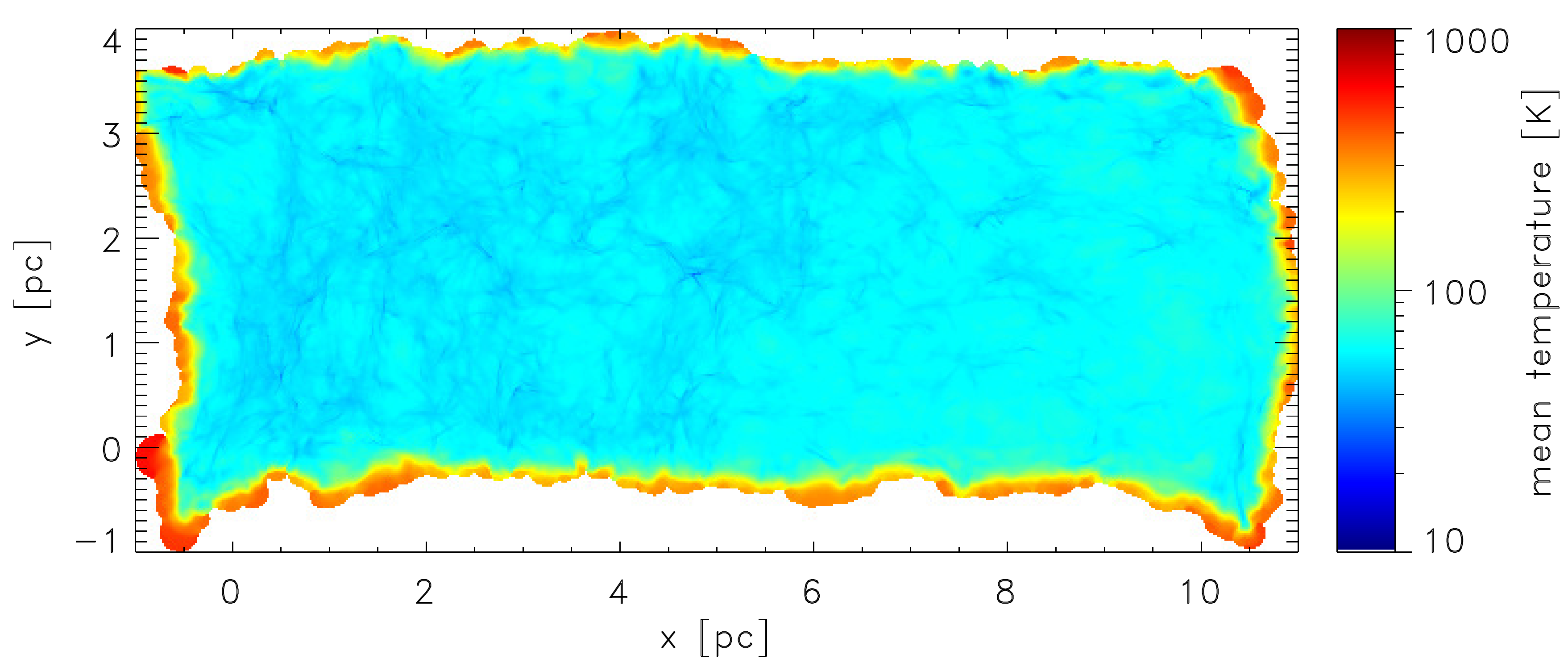}
\includegraphics[width=3.5in]{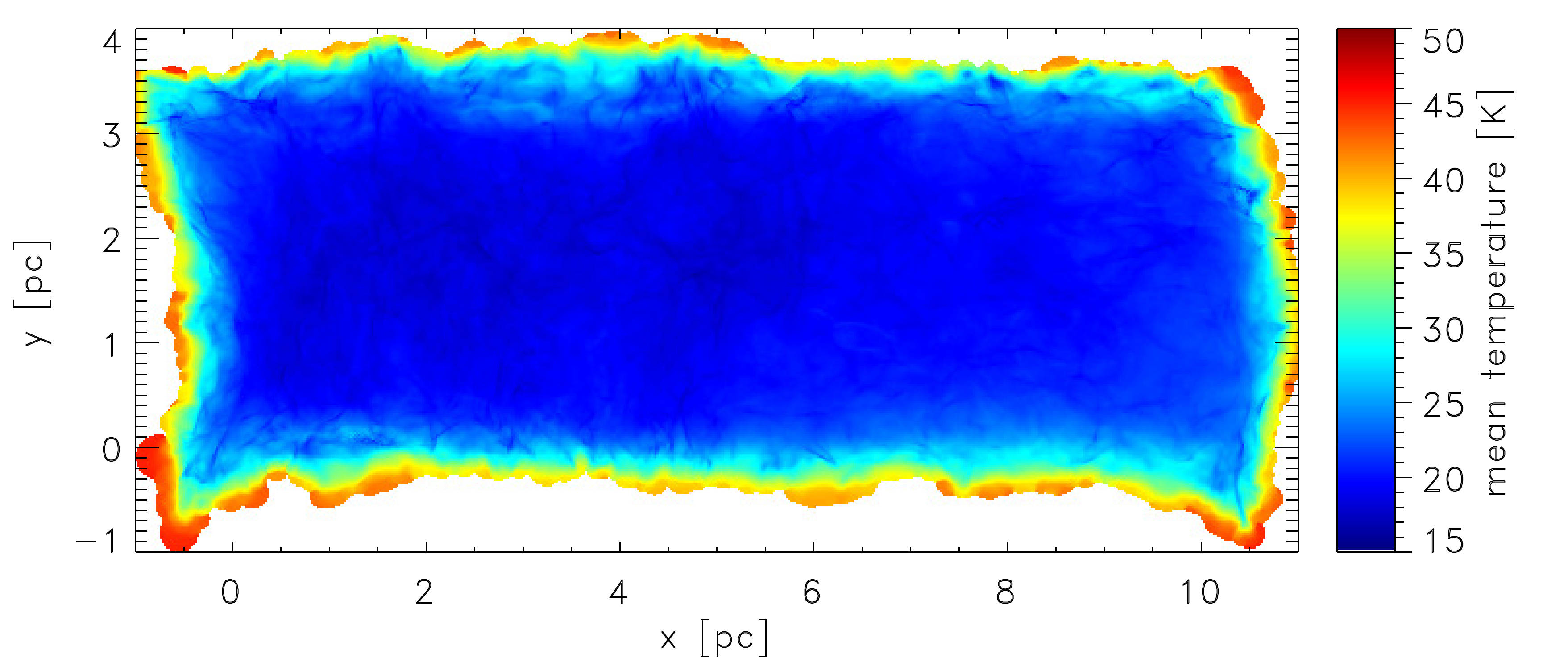}
\caption{\label{fig:pics} Column density, and mean gas and dust temperatures in our fiducial cloud setup (simulation `1' in Table~\ref{table:sims}), with the ISRF set at 1000 \habing, and the CRIR at $3\times \, 10^{-14} \rm s^{-1}$.}
\end{figure}

\section{Computational method}
\label{sec:code}

We perform our simulations using
the smoothed particle hydrodynamics (SPH) code {\sc Gadget2} \citep{springel05}. We have modified the code
to include time-dependent chemistry and a treatment of the main heating and cooling processes (described below). We have also included an implementation of the {\sc TreeCol} algorithm \citep{cgk11}, to obtain column density maps of the sky as seen by each SPH particle. These maps (including total, H$_2$ and CO column densities) are used to calculate the influence of the interstellar radiation field (ISRF) on the gas and the dust.

We assume for simplicity that the spectral shape of the ISRF follows \citet{dr78} in the UV and \citet{bl94} at longer wavelengths. We denote the solar neighbourhood value of the
strength of the ISRF as \habing, and perform simulations with field strengths 100\,\habing~and 1000\,\habing~(see Table~\ref{table:sims}). Note that this multiplicative scaling is done equally at all wavelengths. For our dust model, we use a combination of  values from \citet{oh94} (non-coagulated, thick ice mantle grains) for wavelengths longer than $1 \: \mu{\rm m}$, and from \citet{mmp83} at shorter wavelengths. To compute the visual extinction, we use the relationship $A_{\rm V} = 5.348 \times 10^{-22} (N_{\rm H, tot} / 1 \: {\rm cm^{-2}})$, where $N_{\rm H, tot}$ is the total hydrogen column density  
\citep{bsd78,db96}. 
For simplicity, we do not account for any changes in the extinction curve that may occur due to dust coagulation. For the cosmic-ray ionisation rate (CRIR), we adopt a value of \crlocal $= 3 \times 10^{-17} \rm s^{-1}$ as our solar neighbourhood value \citep{vv00}, and assume that each ionisation event deposits 20\,eV of energy into the gas \citep{gl78}. The dependence of the CRIR on column density is highly uncertain \citep{pado09}, and we assume for simplicity that no attenuation occurs. We do not include the effects of ionization by hard X-rays, as this does not appear to be a major heat source in the Galactic Center, given the relatively low X-ray luminosity \citep{rf04,ssk10}

For the chemistry we adopt the reduced CO network of \citet{nl99}. Details can be found in \citet{gc12b}, and a description of how the chemistry interacts with the ISRF via the {\sc TreeCol} algorithm is given in \citet{gc12a}.

\begin{figure*}
\centerline{
\includegraphics[width=2.35in]{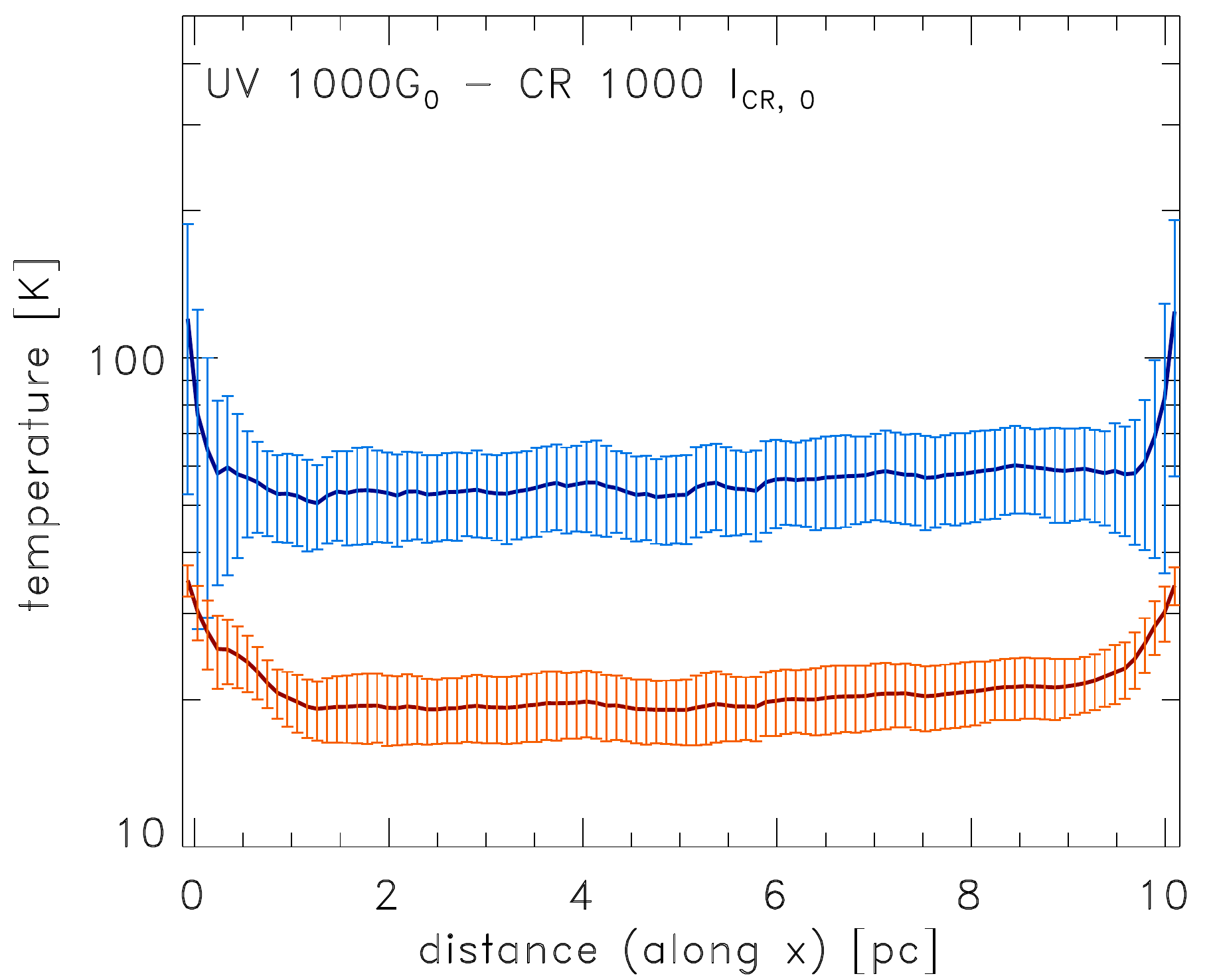}
\includegraphics[width=2.35in]{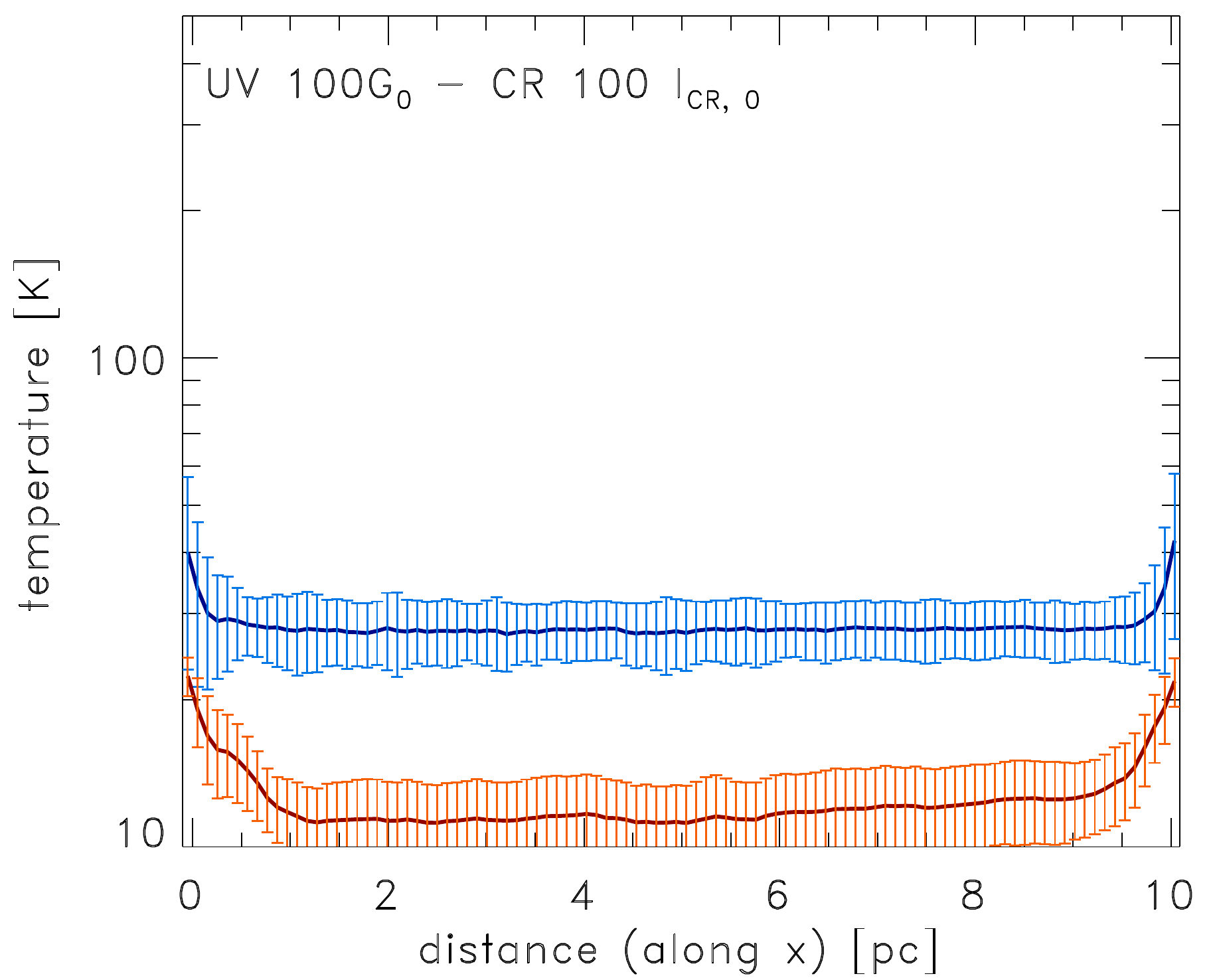}
\includegraphics[width=2.35in]{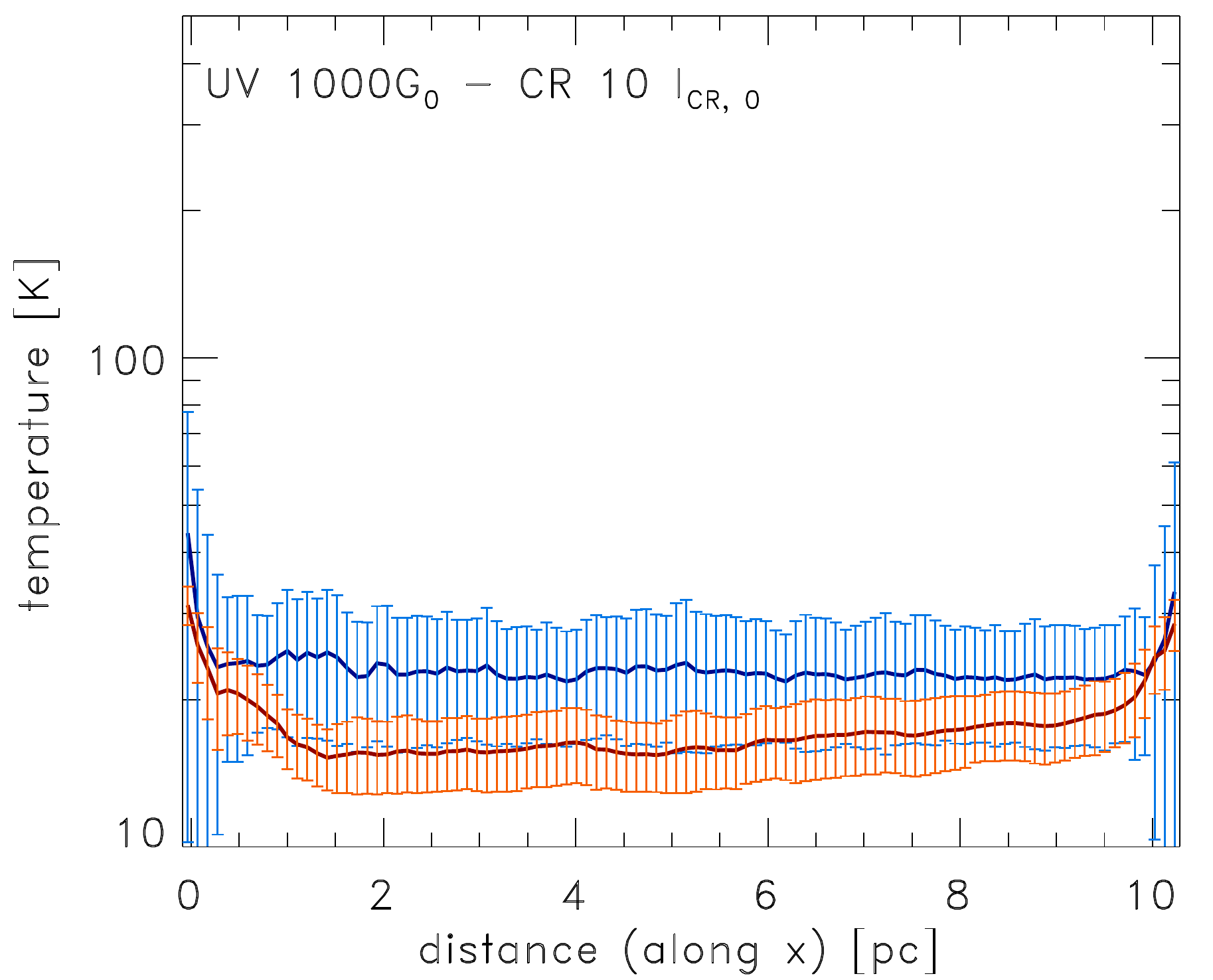}
}
\centerline{
\includegraphics[width=2.35in]{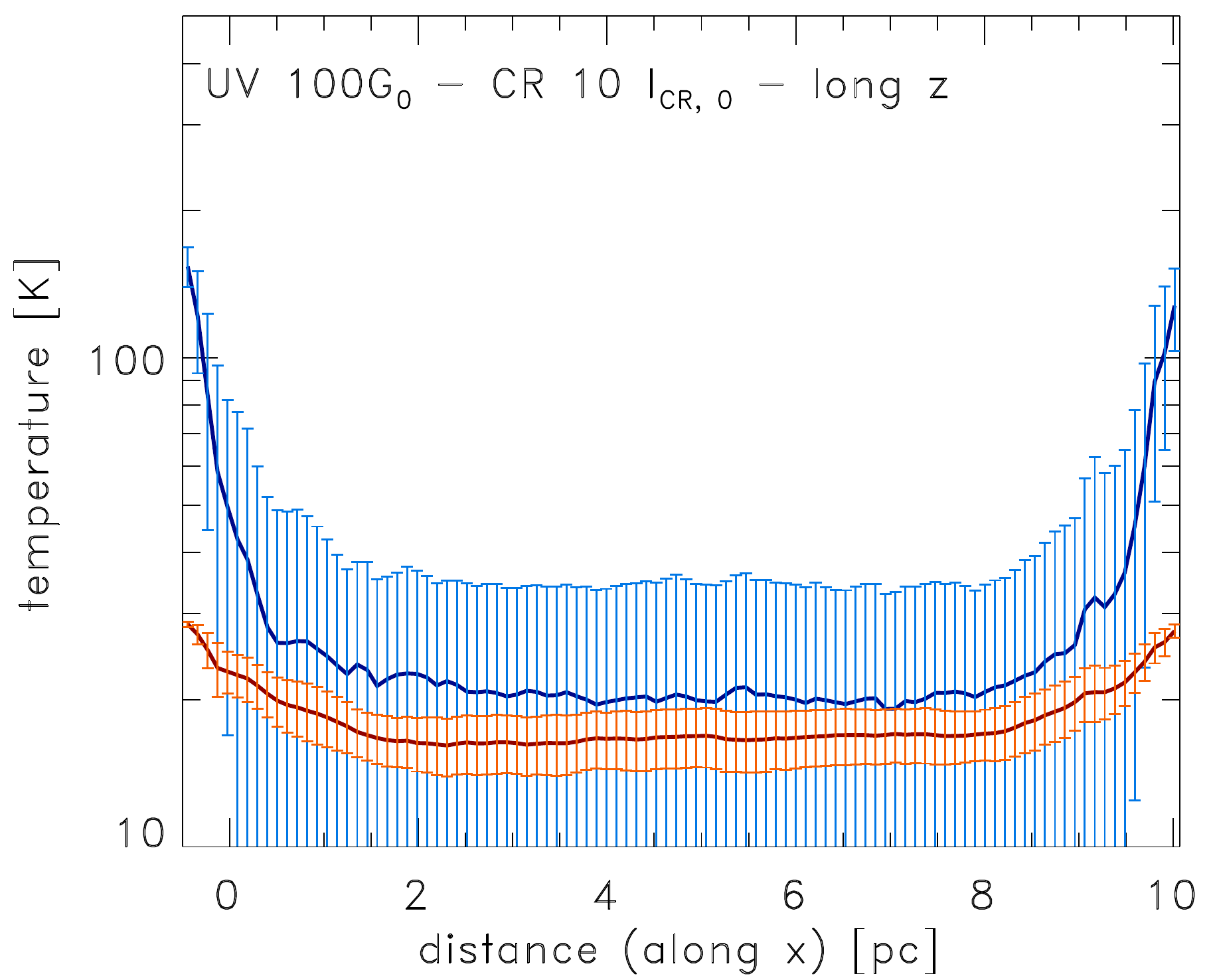}
\includegraphics[width=2.35in]{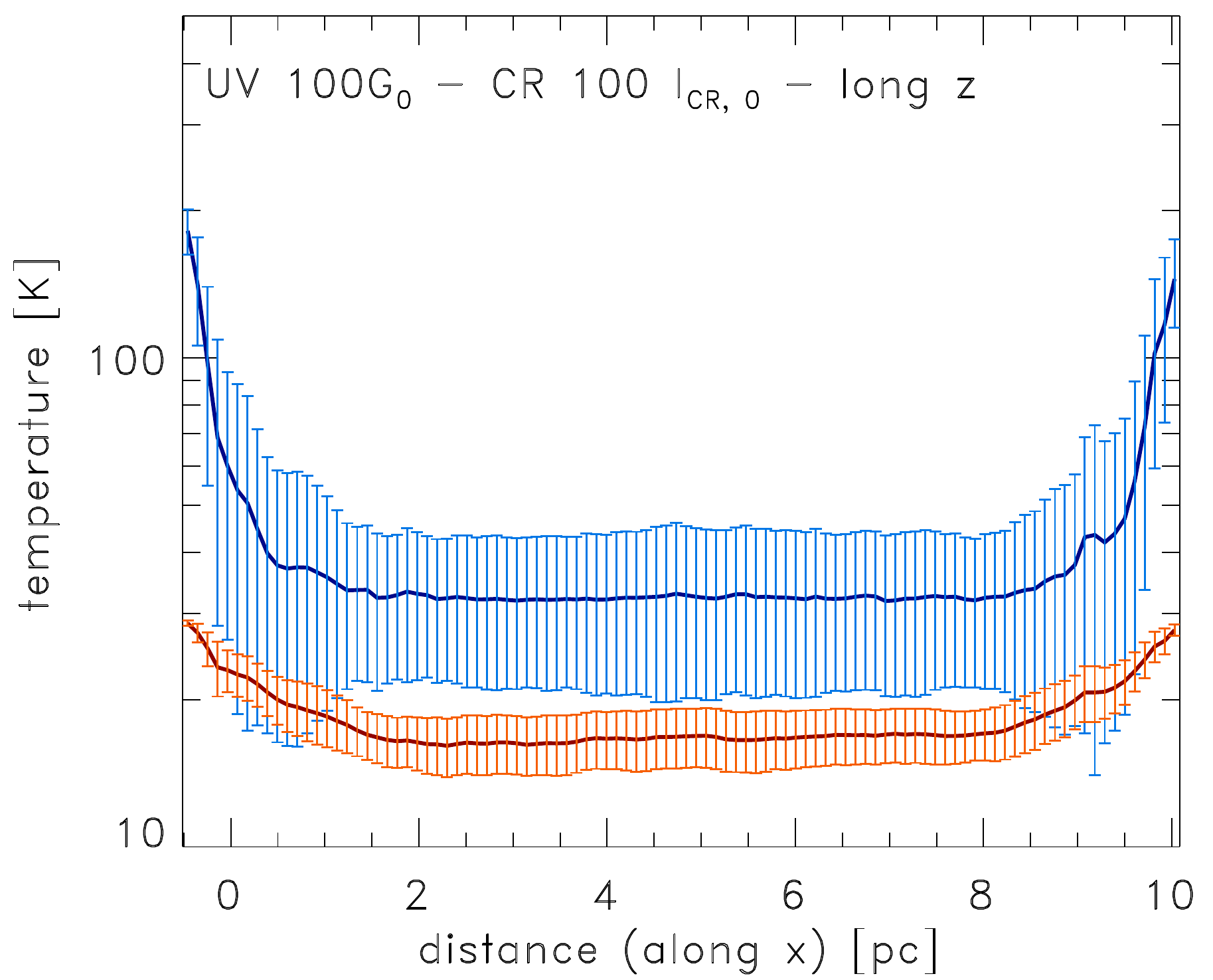}
\includegraphics[width=2.35in]{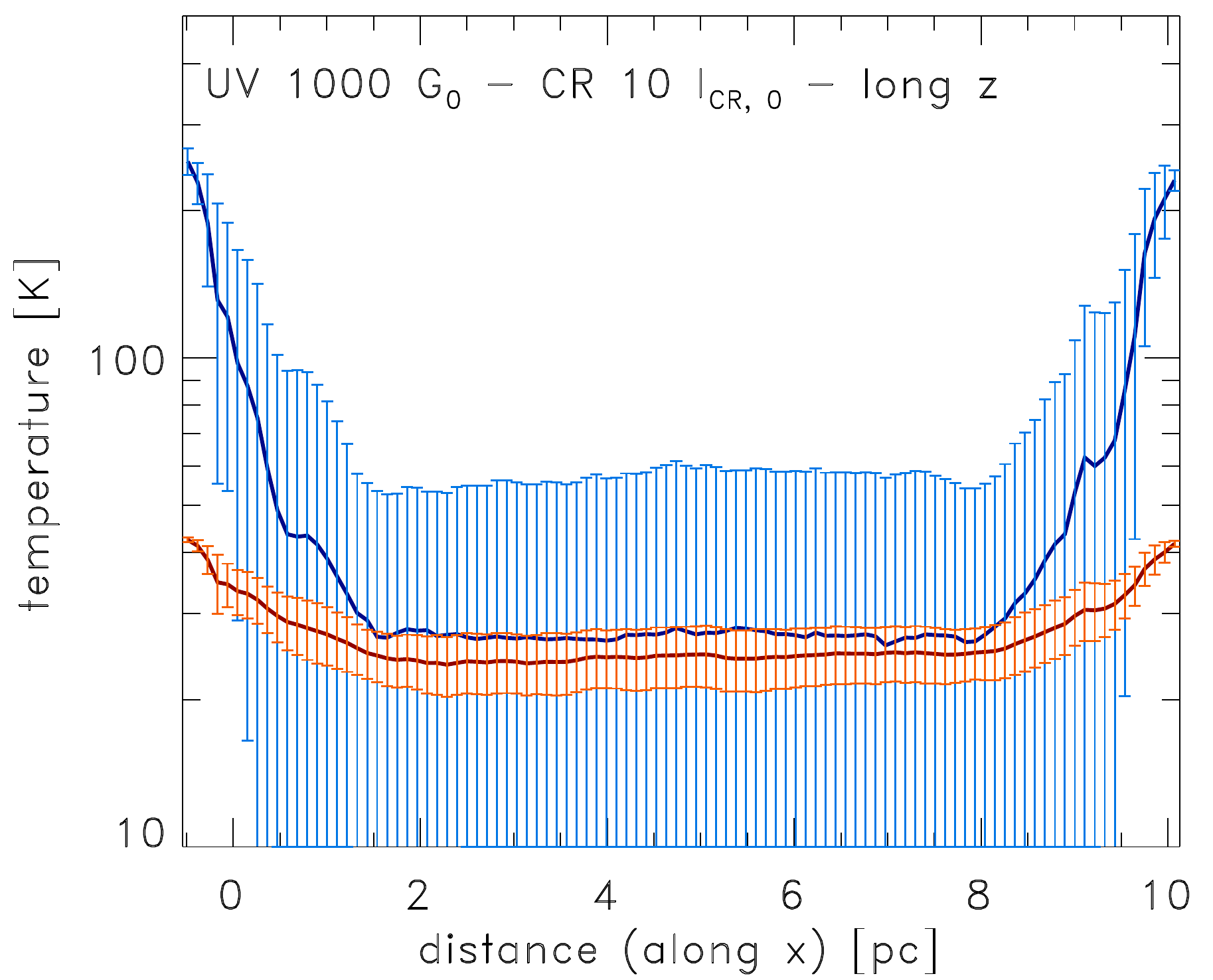}
}
\caption{\label{fig:tprofile} Gas (blue) and dust (red) temperatures as a function of $x$. The top row contains the clouds that have the fiducial setup ($x$ is the longest axis), while the bottom row contains the low density clouds (those with $z$ as the longest axis). The lines denote the mass-averaged temperature along the line of sight. Vertical bars denote the 1-$\sigma$ dispersion.}
\end{figure*}

\section{Initial conditions and model parameters}
\label{sec:sims}

For the initial conditions in this study, we take the cloud properties derived in \citet{Longmore2012} for \thebrick~as a guide: a size of 9.4 pc by 3.4 pc, and a mass of $1.3 \times 10^{5}\,$\solmasp. The clouds are simulated using $2 \times 10^{7}$ SPH particles, and so our mass resolution is $M_{\rm res} = 0.65
\: {\rm M_{\odot}}$ \citep{Hubber2006}. We adopt a simple rectangular cuboid geometry, matching the longer of the two observed dimensions with the $x$-axis in the simulations, and the shorter with the $y$-axis, such that all the clouds have particles placed initially from 0 to 9.4 pc in $x$ ($L_{x}$) and 0 to 3.4 pc in $y$ ($L_{y}$). In the $z$ direction we adopt two values for the extent of the cloud, since the true dimension of \thebrick~along the line-of-sight is unknown. Our first choice is to make the $z$-axis the same length as the $y$-axis, yielding a mean hydrogen nuclei number density $n_{0} =  3.5 \times 10^4 \rm \,cm^{-3}$. This is the setup used in our `fiducial' clouds. Our second choice is to make $z$ the longest axis, with $L_{z} = 17.0 \: \rm pc$. These clouds have an initial density of $6.9 \times 10^3 \rm \,cm^{-3}$ and are our `low-density' clouds. All the clouds are given non-thermal support in the form of a turbulent velocity field, which has a power spectrum $P(k) \propto k^{-4}$. The turbulence is permitted to decay as the cloud evolves. We fix the initial 3D turbulent velocity dispersion based on the observational data: \citet{Longmore2012} report a linewidth of 15.1 $\rm km\, s^{-1}$ for \thebrick, equivalent to a 1D velocity dispersion of 6.4 $\rm km\, s^{-1}$, and hence to a 3D velocity dispersion of 11.12 $\rm km\, s^{-1}$, assuming isotropic turbulence.

We perform three simulations for each of our two cloud models, varying the strength of the ISRF and the magnitude of the CRIR.  An overview of the simulations can be found in Table~\ref{table:sims}.  A central assumption here is that the shape of the radiation field and the cosmic ray energy spectrum are the same locally and in the Galactic Centre, and that it is only the normalization of each that changes.

In view of the high densities probed by our initial conditions, we assume that the hydrogen in our clouds starts as H$_{2}$. However, we start with the carbon in the form of C$^+$, and allow it to self-consistently evolve to form C and CO. As we discuss in \S \ref{sec:thermo}, the clouds are already in chemical equilibrium at the point at which we perform our analysis.

\section{Gas and Dust Temperature}
\label{sec:temp}

\begin{figure}
\includegraphics[width=3.5in]{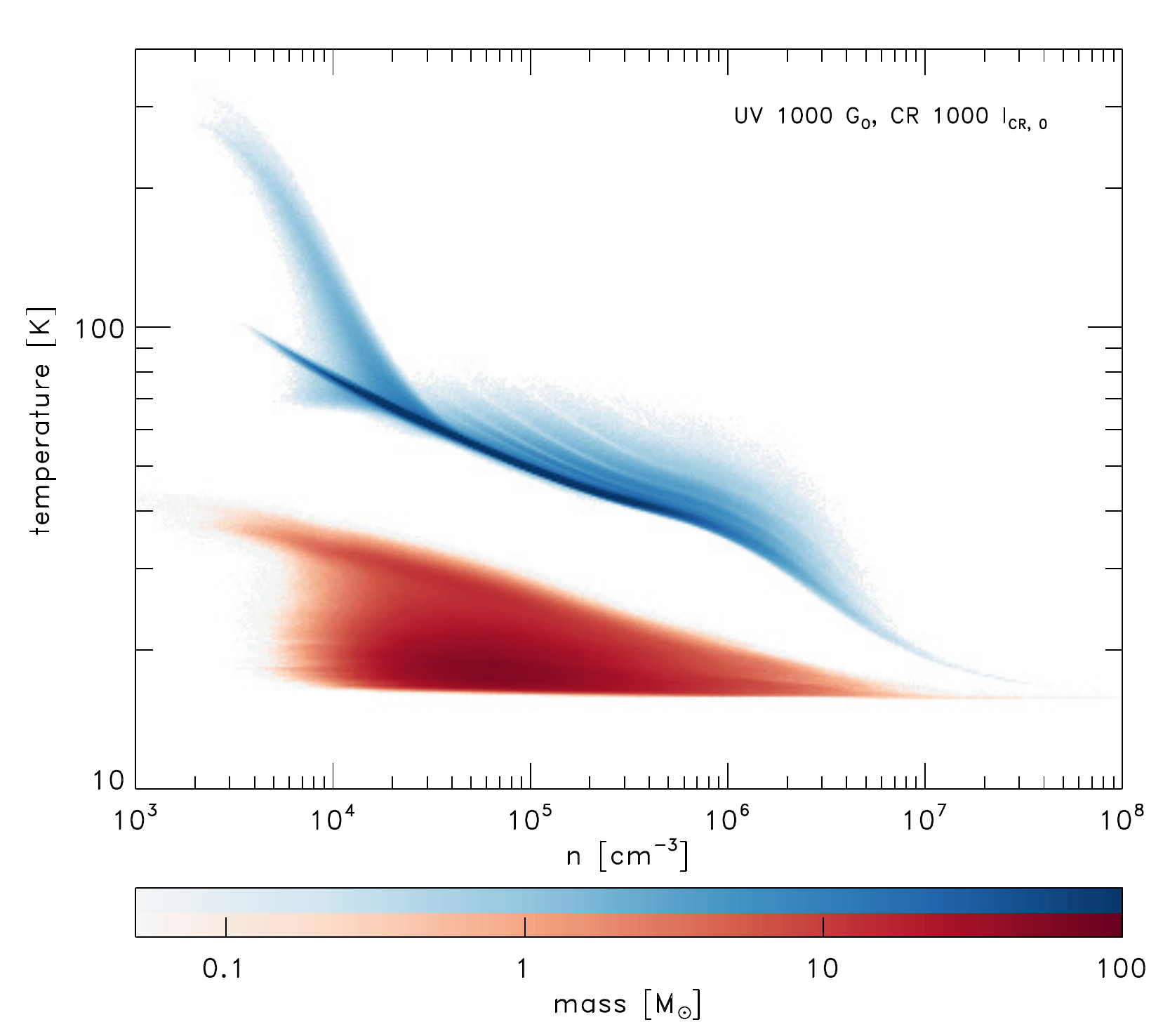}
\caption{\label{fig:phase} Gas (blue) and dust (red) temperatures as a function of density in our fiducial cloud (model `1' in Table \ref{table:sims}).}
\end{figure}

Using {\em Herschel} observations, \citet{Longmore2012} show that the dust temperature varies smoothly from 19\,K in the cloud centre to 27\,K at the edge. Observational constraints on the gas temperature of \thebrick~have existed for some time: \citet{guesten1981} derive rotation temperatures of $\sim$45\,K using ammonia transitions, corresponding to an average kinetic temperatures of $\sim$80\,K \citep{walmsleyung83}.  A recent formaldehyde survey \citep{ao2013} finds average kinetic temperatures of 65--70\,K, which agrees within the uncertainties. However, observations of high-excitation ammonia lines suggest that \thebrick~has a complex gas temperature structure, with components up to 400\,K, that has yet to be modeled (E. Mills, 2013, private communication).  What environmental conditions are required to produce such temperatures?

The typical features of our cloud are illustrated in Fig. \ref{fig:pics}, which shows the column densities of one of the clouds in the $x$-$y$ plane (i.e.\ integrating along $z$), and the accompanying mean gas and dust temperature maps. This cloud is our most extreme case studied, with ${\rm I_{ISRF}} = 1000 \, {\rm G}_{0}$, and ${\rm I_{CR}} = 1000\, {\rm I_{CR, 0}}$, and our `fiducial' cloud geometry. However, the features of this cloud are mirrored in our other simulations -- the clouds have a hot skin and a relatively cool interior, and are highly structured by the supersonic turbulence. The images in Fig.~\ref{fig:pics} are taken just as the first collapsing core exceeds a density of around $10^8 \: \rm cm^{-3}$, and so represent the state of the cloud at the onset of star formation. All the other clouds in this study will be presented at the same point in their evolution.

In Figure~\ref{fig:tprofile}, we show the gas and dust temperatures in the clouds as a function of the position along the $x$-axis. The most obvious feature of these profiles is that the gas and dust have different temperatures throughout the cloud. They are not thermodynamically coupled on the scales shown here, consistent with the observations mentioned above.

The profiles also reveal how the environment affects the cloud temperature. We see that the cosmic rays are responsible for heating the gas, while the ISRF is primarily responsible for heating the dust. Such a result is expected. The high column density of this cloud means that photo-electric emission in the cloud interior is strongly suppressed, as the UV photons responsible for it are readily absorbed near the surface of the cloud. As such, the ISRF can play only a minor role in directly heating the gas. On the the other hand, as the cosmic rays have no attenuation in our model, they are free to heat the cloud's gaseous interior throughout. The ISRF can, however, heat the dust at the centre of the cloud, as this heating comes primarily from longer wavelength photons, which are able to penetrate much further than the UV photons. In summary, for clouds with such an extreme column density as \thebrick, the heating of the dust and gas is effectively split into two components.

Our 3D modelling results suggest that for our fiducial cloud model, the environmental parameters that best reproduce the observed temperatures are \uvrad~$=$ 1000  \habing, and \crrate = 1000 \crlocal. Reducing either of these values by a factor of ten results in gas or dust temperatures that are too low to agree with the observations.

One potential source of error is simply that we have underestimated the extent of \thebrick~along the observed line-of-sight, and so the true effective column of the cloud is much smaller than we are assuming in the fiducial models. However we find that similar environmental conditions are also required when we consider our lower-density version of \thebrick. These models are shown on the bottom row of Fig \ref{fig:tprofile}. Even in these lower column density clouds, we see that the ISRF is mainly responsible for determining the dust temperatures (i.e.\ there is very little gas-dust thermodynamic coupling), and the CRIR is mainly responsible for determining the gas temperatures. Our dust temperatures are now a little higher than the observed values throughout the cloud, suggesting that for this geometry the \uvrad~would need to be lower than 1000~\habing. However we see that by 100~\habing, the ISRF is already too low to explain the observed temperatures. Also, we see that \crrate = 100 \crlocal~ results in a gas temperature of around 30~K in the interior of the cloud -- again, this is inconsistent with the observations.

Figure~\ref{fig:tprofile} also shows that the geometry of the cloud affects the temperature gradients along the cloud. This is particularly evident when one looks at the gas temperature, especially when \uvrad~is high (see e.g.\ the bottom right panel). This implies that it should be possible to constrain both the total ISRF and the cloud's geometry by fitting the gradient of the gas temperature in the cloud modelling. Such a study is worth revisiting once maps of the gas temperature with sub-parsec resolution become available.

Finally, we note that both the gas and dust temperatures can vary considerably along a line of sight from the averages shown in Fig. \ref{fig:tprofile}. This can already be seen in the images in Fig. \ref{fig:pics}. However, we also show in Fig. \ref{fig:phase} how the temperatures vary as a function of density in our fiducial case. We see that at high densities ($> 10^6 \: \rm cm^{-3}$), once the dust and gas thermally couple, the temperatures can be relatively cold.

\begin{figure}
\includegraphics[width=3.5in]{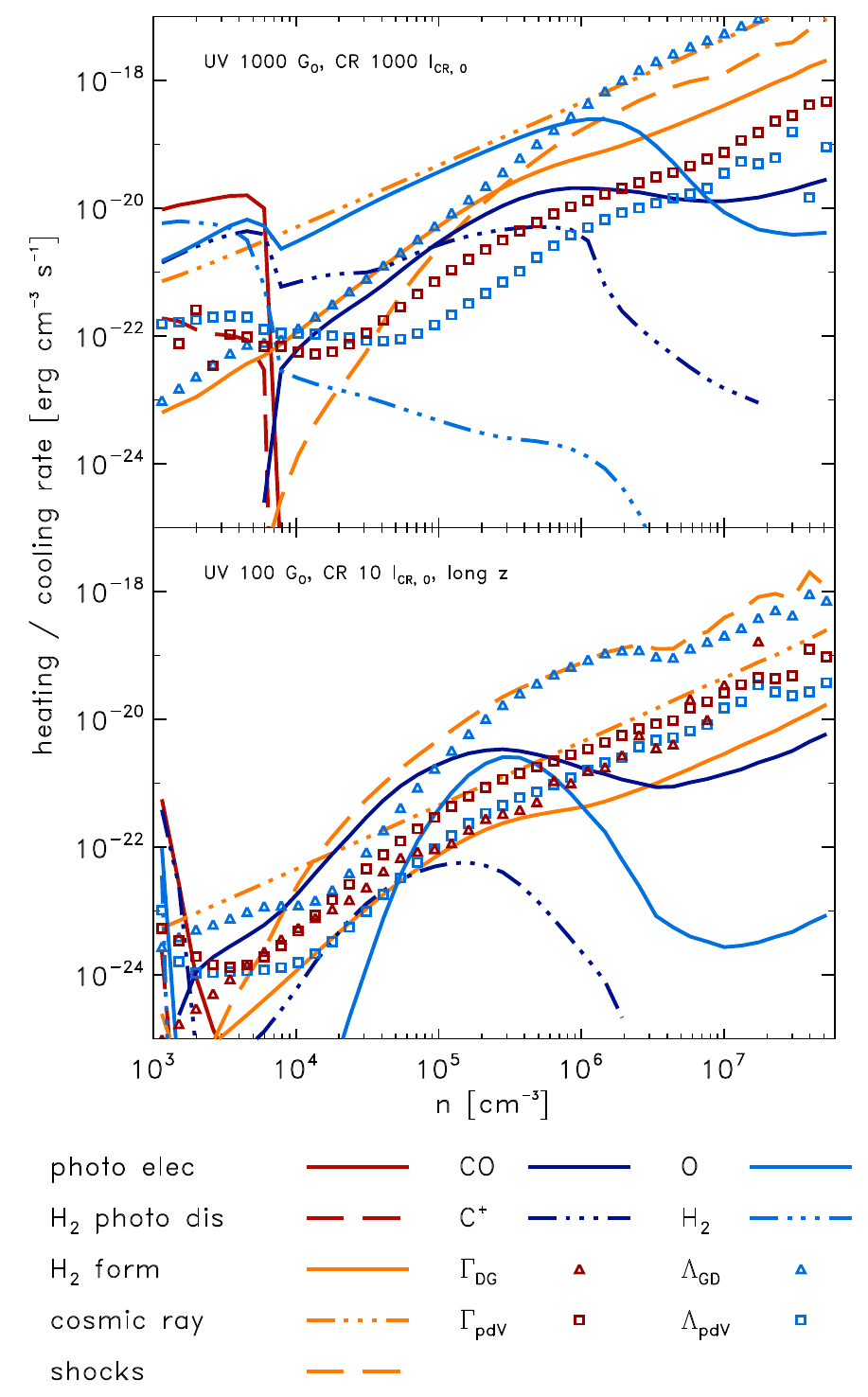}
\caption{\label{fig:therm} Processes responsible for heating and cooling the gas in two very different cloud models (clouds 1 and 4 from Table \ref{table:sims}). Heating processes are shown in red and orange and cooling processes are represented in blue. Two processes -- $p {\rm d}V$ work and gas-dust thermal coupling -- can produce either heating or cooling depending on the circumstances. Heating and cooling associated with compression and expansion are denoted by $\Gamma_{p{\rm d}V}$ and $\Lambda_{p{\rm d}V}$, respectively, while the transfer of energy from the gas to the dust is denoted by $\Lambda_{\rm GD}$ and that from the dust to the gas by $\Gamma_{\rm DG}$. The plotted quantities represent the median values at each density.}
\end{figure}

\section{Heating and Cooling processes}
\label{sec:thermo}

In this section we investigate the heating and cooling processes for the gas in more detail.  The dominant processes that govern the gas temperature are shown as functions of density in Fig.~\ref{fig:therm} for the two most extreme cases: our fiducial cloud (n$_0 = 3.5 \times 10^4\,\rm cm^{-3}$) with ${\rm I_{ISRF} = 1000 \, G_{0}}$ and ${\rm I_{CR} = 1000 \,I_{CR, 0}}$,  and one of the lower-density clouds (n$_0 = 6.7 \times 10^3\,\rm cm^{-3}$), with ${\rm I_{ISRF} = 100 \, G_{0}}$ and ${\rm I_{CR} = 10 \,I_{CR, 0}}$. 

In both clouds, the dominant heating processes follow a broadly similar pattern. At the lowest densities, which represent the outskirts of the clouds in these simulations, the dominant heat source is photoelectric emission from dust grains. This falls off sharply as we move to higher densities as a result of the increasing extinction as one moves into the cloud's interior. At slightly higher densities, the heating caused by cosmic rays starts to dominate the thermal balance. In the case of the hotter, denser cloud, this process remains the main heating source until we reach a number density $n = 10^8 \: {\rm cm^{-3}}$, corresponding to our resolution limit. In the lower density cloud, embedded in the less extreme environment, shock heating becomes the main source of heat input to the gas at densities above $n \sim 10^{5} \: {\rm cm^{-3}}$. Note that since neither compression nor shock-heating are dominant in the high CRIR case, the temperature of the cloud cannot be used to determine its age.

When we compare the main cooling processes, we also find some similarities. In the low-density outskirts, where the gas is warm and there is little CO, we find that C$^+$ and neutral oxygen emission are the main coolants, as in the low-density ISM. Given the high densities and temperatures of the cloud's skin, and the fact that we start with the hydrogen in molecular form, we also find that H$_2$ can be an effective coolant at the outskirts. 

As we move into the cloud, however, the gas temperature drops and the C$^{+}$ recombines to form C and then CO. The identity of the dominant coolant therefore changes. In the low-density cloud, CO cooling dominates in this slightly denser regime, just as is the case in local molecular clouds. In the denser cloud model, however, CO never dominates; instead, atomic oxygen becomes the main coolant. This difference in behaviour is a result of the CRIR in these two clouds. In the higher density cloud, the much higher CRIR creates many He$^{+}$ ions that react destructively with the CO molecules:
\begin{equation}
{\rm CO} + {\rm He^{+}} \rightarrow {\rm C^{+}} + {\rm O} + {\rm He}.
\end{equation}
It also keeps the gas warm enough to excite the fine structure lines of atomic oxygen. In the lower density cloud with the much lower CRIR, both of these effects are less important, and hence atomic oxygen never becomes the dominant coolant.  Since we need a large CRIR to explain the observed gas temperatures, the implication is that the cooling of gas in \thebrick~(and probably also in other Galactic Centre clouds) is dominated over a significant range in densities by emission from atomic oxygen.

At very high densities, dust becomes the most effective source of cooling. However this does not occur until the gas density is more than an order of magnitude higher than the mean cloud density, and hence we expect that $T_{\rm gas} = T_{\rm dust}$ only in the densest gas within \thebrick, with most of the volume of the cloud having $T_{\rm gas} \neq T_{\rm dust}$. As already noted, this expectation is supported by the available observational data on the gas and dust temperatures.

The effect of the clouds' environment on the chemical balance is summarised in Table \ref{table:sims}. We see that strong ISRFs and CRIRs have little effect on the H$_2$ fraction, and so we would expect the true molecular state of the cloud to be relatively independent of the environment. However, the CO fraction varies by around an order of magnitude in the models, implying that its ability to trace the molecular state of the gas is a strong function of the environment. Since the clouds initially have all of their carbon in the form of C$^+$, one might argue that we have simply ended our simulations too soon to pick up all of the CO. However, we see that in the clouds with smaller CRIRs over half of the carbon is  in CO, suggesting that there is sufficient time available for it to form in large quantities. As such, the low CO abundances in the clouds with high CRIR are due to real differences in their chemical evolution.

\section{Discussion}
\label{sec:sum}

Our results suggest that the CRIR and ISRF around \thebrick~should be 1000 times the solar neighbourhood values, in order to obtain temperatures consistent with the values derived from observations.  Such radiation and CR fields could be produced by enhanced star formation activity, higher stellar densities, or some combination of both.  \citet{yz09} measured the star formation rate (SFR) in the GC to be ~50--100 times the local SFR.  If the CRIR and ISRF are set solely by star formation, our results suggests that the local SFR near \thebrick~is about an order of magnitude higher than the mean SFR of the central molecular zone \citep{MorrisSerabyn1996, yz09}. 

Similarly, the CRIR that we require is significantly higher than the values found for local dense clouds. However, there is considerable observational evidence that the ionization rate is higher in the Galactic Centre. For example, \citet{oka05} estimate a value of 2--7$ \times 10^{-15} \: {\rm s^{-1}}$ in diffuse gas along several GC sightlines,
while \citet{yz07} infer a value of 2--50$ \times 10^{-14} \: {\rm s^{-1}}$ within GC clouds, based on observations of the fluorescent 6.4~keV K$\alpha$ iron line. Our required value of a few times $10^{-14} \: {\rm s^{-1}}$ is compatible with these values, given the large uncertainties.

Our models also suggest that the neutral oxygen emission coming from \thebrick~should be significantly higher than that seen typical molecular clouds. This could provide an independent test of the models presented in this paper.

\section{acknowledgements}
The authors would like to thank Katharine Johnston and Elizabeth Mills for their enlightening discussions on \thebrick. We acknowledge financial support from the DFG via SFB 811 ``The Milky Way System'' (sub-projects B1 and B2), and from the Baden-W\"{u}rttemberg-Stiftung by contract research via the programme Internationale Spitzenforschung II (grant P- LS-SPII/18). PCC and SER are supported by grant CL 463/2-1 and RA 2158/1-1, respectively, which are part of the DFG-SPP 1573. The simulations presented in this paper were performed on the {\em Milkyway} supercomputer at the J\"ulich Forschungszentrum, funded via SFB 811.

\begin{acknowledgments}

\end{acknowledgments}

\end{document}